\documentclass[10pt,letterpaper,twocolumn]{article}

\usepackage[
    letterpaper,
    top=0.65in,
    bottom=0.55in,
    left=0.65in,
    right=0.65in,
    columnsep=0.24in
]{geometry}

\usepackage{times}
\usepackage{mathptmx}

\usepackage{amsmath}
\usepackage{amssymb}

\usepackage{graphicx}
\usepackage{booktabs}
\usepackage{array}
\usepackage{multirow}
\usepackage{caption}
\usepackage{cuted}

\usepackage{url}
\usepackage{microtype}
\usepackage{fancyhdr}
\usepackage{titlesec}
\usepackage{enumitem}
\usepackage{float}

\makeatletter
\renewenvironment{thebibliography}[1]
  {%
    \list{\@biblabel{\@arabic\c@enumiv}}%
    {%
      \settowidth\labelwidth{\@biblabel{#1}}%
      \leftmargin\labelwidth
      \advance\leftmargin\labelsep
      \@openbib@code
      \usecounter{enumiv}%
      \let\p@enumiv\@empty
      \renewcommand\theenumiv{\@arabic\c@enumiv}%
    }%
    \sloppy
    \clubpenalty4000
    \@clubpenalty\clubpenalty
    \widowpenalty4000%
    \sfcode`\.\@m
  }
  {%
    \def\@noitemerr
      {\@latex@warning{Empty `thebibliography' environment}}%
    \endlist
  }
\makeatother

\setlist{
    topsep=2pt,
    itemsep=1pt,
    parsep=0pt,
    partopsep=0pt
}

\titleformat{\section}
  {\normalfont\bfseries\centering}
  {\Roman{section}.}
  {0.35em}
  {}
  [.]

\titlespacing*{\section}
  {0pt}
  {8pt plus 2pt minus 1pt}
  {5pt plus 1pt minus 1pt}

\titleformat{\subsection}
  {\normalfont\bfseries}
  {\Alph{subsection}.}
  {0.35em}
  {}

\titlespacing*{\subsection}
  {0pt}
  {7pt plus 1pt minus 1pt}
  {4pt plus 1pt minus 1pt}

\begin{document}


\twocolumn[
\begin{center}


\vspace*{0.04in}

{\fontsize{14}{16}\selectfont
\fontfamily{ptm}\selectfont
\bfseries
TATVA: A Reinforcement Learning Framework for Quantum Circuit Synthesis
}

\vspace{0.055in}


{\fontsize{10}{11.5}\selectfont
\fontfamily{ptm}\selectfont
Om Bhamare$^{1}$, Aryan Jadhav$^{1}$, Manas Shinde$^{1}$,
Prithvi Shinde$^{1}$
}

\vspace{0.025in}


{\fontsize{9}{10.5}\selectfont
\fontfamily{ptm}\selectfont
$^{1}$K.K.\ Wagh Institute of Engineering Education and Research
}

\vspace{0.035in}


{\fontsize{9}{10.5}\selectfont
\fontfamily{ptm}\selectfont
\begin{tabular}{ccc}
Dr. Uday Wad$^{2}$ &
Dr. Snehal Kamalapur$^{1}$ &
Dr. S. S. Bhandare$^{1}$ \\[1.5pt]
Professor of Practice &
HOD, Department of Computer Engineering &
Assistant Professor
\end{tabular}
}

\vspace{0.055in}

\end{center}


\begin{center}

{\fontsize{10}{11.5}\selectfont
\fontfamily{ptm}\selectfont
\begin{minipage}{0.94\textwidth}

\setlength{\parindent}{0pt}
\setlength{\parskip}{0pt}

One of the most important initial steps in quantum computing is High-fidelity Quantum State Preparation, because the errors in it can affect the final computational result. Therefore one of the major challenges is to design a quantum circuit that will achieve the desired target state accurately with smaller number of gates. Traditionally the circuits are designed using some predefined rules and mathematical decompositions but it makes it difficult to design for different and hard target states. This paper presents TATVA, a reinforcement learning (RL) system for synthesizing quantum circuits. TATVA views the circuit synthesis as a sequential decision problem in which the agents select one gate at a time and calculate the fidelity for each applied gate and the gate with the highest fidelity is finally selected for the circuit. The system introduces a novel parallel architecture that uses two RL agents, Deep Q-Network (DQN) and Proximal Policy Optimization (PPO) along with Qiskit's state vector simulator. The system can synthesize upto 5 qubits while achieving a fidelity of 0.999999 and having a compact circuit. After achieving the desired fidelity, it performs a post-hoc optimization step which reduces the circuit depth while preserving the fidelity obtained. The circuit is finalized using the fidelity, success rate, gate count and circuit depth. The target states that are not used during the training are used to test the system to check if the system has learned properly or not. This approach aims automatic quantum circuit synthesis with high fidelity using reinforcement learning.

\end{minipage}
}

\vspace{0.035in}


{\fontsize{10}{11.5}\selectfont
\fontfamily{ptm}\selectfont
\begin{minipage}{0.94\textwidth}

\setlength{\parindent}{0pt}
\setlength{\parskip}{0pt}


\end{minipage}
}

\vspace{0.035in}

\end{center}
]


\section{INTRODUCTION}

Quantum computing uses a different approach from classical computing.
It uses properties like superposition and entanglement to process the
information. In circuit models, the initial state is turned into
desired state by applying quantum gates on the qubit. In this process,
an important initial step is Quantum State Preparation (QSP).

Preparing a high-fidelity state is very important because errors in it can affect the final result of the quantum computation
\cite{liu}. QSP is used in applications of quantum machine learning (QML),
quantum simulation and preparing quantum state in different quantum
algorithms \cite{liu}.

Preparing a reliable quantum state is thus very important, and this
leads to the problem of quantum circuit synthesis: how can we
automatically find a quantum circuit sequence that produces the desired
quantum state? The evaluation of the resultant circuit and its
quality is measured not only using fidelity but also by its number of gates
and circuit depth.

Compact and smaller circuit synthesis is very important because
larger circuits increase the possibility of errors \cite{fosel}. Thus it
becomes difficult for different circuit structures and problems \cite{fosel}.
Thus previous researchers have done research and have explored
reinforcement learning for automatically learning quantum circuit
optimization \cite{fosel,zhang,moro,keren}.

Reinforcement Learning (RL) allows the agent to learn through continuous
trial and error in the environment. The agent analyzes the current state,
applies the gate, receives the reward, and based on it decides the result
\cite{zhang,liu,mnih}. Thus the current quantum state can be viewed as an
environment state, gate selection as an action, and the resulting fidelity
as a reward \cite{zhang,liu}.

In previous studies, deep reinforcement learning has been used for
quantum circuit optimization \cite{fosel,liu,keren}. To solve this problem,
\textbf{TATVA} uses DQN and PPO agents running in parallel with a custom
reinforcement learning environment, Qiskit's state vector simulator
and a fidelity based reward function to synthesize quantum circuit
from 1--5 qubit target states.

The resulting quantum circuits are optimized further in the final
step while maintaining the fidelity to reduce the number of
unnecessary gates.


\section{QUANTUM CIRCUIT SYNTHESIS AND REINFORCEMENT LEARNING}

\subsection{Representing a Quantum State}

An $n$-qubit system is described by a statevector of $2^n$ complex
amplitudes. TATVA takes a target statevector and its associated
qubit count as input; for the input to be valid, its dimension must
satisfy

\begin{equation}
\text{Dimension} = 2^n,
\end{equation}

and its amplitudes must obey the normalization condition

\begin{equation}
\sum_i |\alpha_i|^2 = 1.
\end{equation}

Once validation is done, the target statevector is represented as a
complex128 NumPy array and wrapped in Qiskit's statevector object.

\subsection{Building the Circuit}

Every episode starts with $n$ qubits initialized to

\begin{equation}
|\psi_0\rangle = |0\rangle^{\otimes n}.
\end{equation}

At every timestep, the agent appends one gate, selected from the
discrete action set

\begin{equation}
\{H,\,X,\,Y,\,Z,\,\text{CNOT},\,R_X,\,R_Y,\,R_Z\}.
\end{equation}

The selected gate is applied to the appropriate qubit, and the
circuit is again simulated in Qiskit to obtain the resulting state.

\subsection{Measuring Fidelity}

The similarity of the synthesized state to the target is quantified by
quantum state fidelity,

\begin{equation}
F = |\langle \psi_{\text{target}} | \psi_{\text{current}} \rangle|^2,
\end{equation}

and TATVA treats

\begin{equation}
F > 0.999999
\end{equation}

as the threshold a circuit must clear before it counts as a
successful synthesis.

\subsection{Shaping the Reward}

Fidelity by itself is an incomplete objective - a circuit that
reaches the target using forty gates is a considerably worse outcome than one
that reaches it with five, even though both may report the same fidelity \cite{fosel,zhang}. The reward function is built to reflect that trade-off \cite{zhang}:

\begin{equation}
r = F - \lambda \times \text{Gate Count},
\end{equation}

where $\lambda$ penalizes each additional gate, and a further bonus
is granted once fidelity exceeds 0.999999. This pushes the agent toward
circuits that are accurate \emph{and} economical, rather than accurate at any gate-count cost.


\section{PROPOSED TATVA METHODOLOGY}

TATVA is organized into five stages that depend on one another:
input processing, RL-based gate selection, quantum simulation,
circuit optimization, and output generation. Fig.~\ref{fig:system}
lays out the full pipeline, including the configuration and storage
components -- the replay buffer, model checkpoints, and training
logs -- that support the two learning agents described in this
section.


\subsection{Preparing the Target State}

The pipeline begins by taking in the user-supplied target
statevector and qubit count, checking that the dimensionality
conforms to $2^n$, and normalizing the vector under the L2 norm.
The normalized vector is converted to a complex128 array and
represented in Qiskit as a Statevector object, producing a
validated target for the RL environment to train against, as shown
in Fig.~\ref{fig:target}.


\subsection{The RL Environment}

The environment initializes the specified number of qubits to
$|0\rangle$ with an empty circuit. At each timestep, the agent
observes the circuit's current state and chooses a gate which is suitable, which
Qiskit applies through simulation; fidelity to the target is
recalculated, and the reward follows from this fidelity together with
the gate-count penalty and, where it applies, the success bonus. An
episode ends when the fidelity threshold is reached or when
the maximum gate count is used up, whichever happens first. Fidelity,
gate count, and episode outcome are all logged for later
evaluation, following the loop shown in Fig.~\ref{fig:rlenv}.


\subsection{DQN Agent}

The first of TATVA's two agents is a \textbf{Deep Q-Network} \cite{mnih}. The
current environment state is fed to the Q-network, which outputs an
estimated value for each available action; the agent follows an
epsilon-greedy policy, balancing exploitation of its present
estimate against exploration of the alternatives \cite{mnih}. Each transition

\begin{equation}
(s, a, r, s')
\end{equation}

is stored in a replay buffer \cite{mnih}, and mini-batches sampled from that
buffer are used during training to calculate the temporal-difference
target and update the network. A separate target network,
synchronized with the online network at fixed intervals, keeps
training stable by damping oscillations in the learned value
estimates \cite{mnih}.


\subsection{PPO Agent}

The second agent is \textbf{Proximal Policy Optimization} \cite{schulman}. An actor
network maps the observed state onto a probability distribution
over the available gates; the sampled action is executed and the
resulting trajectory is stored. A critic network estimates the
value of the observed state, and Generalized Advantage Estimation
(GAE) turns those value estimates into an advantage signal. The
policy is then updated through a clipped surrogate objective with
an entropy bonus attached to preserve exploration \cite{schulman}, and both the
actor and critic parameters are updated by gradient descent.

Fig.~\ref{fig:dqnppo} sets the two agents side by side: the DQN
branch follows epsilon-greedy action selection with replay-buffer
training, while the PPO branch follows actor-critic rollout, GAE,
and the clipped-objective update, and both branches return an
updated policy to the RL environment.


\subsection{Simulation and Fidelity Check}

The gate sequence the agent has produced so far is assembled into a
quantum circuit and run on Qiskit's statevector simulator. The
resulting state is compared to the target using the fidelity
expression defined in Section II. If $F > 0.999999$, the synthesis
criterion has been met; otherwise the reward is computed and training
continues until the gate budget runs out, as shown in
Fig.~\ref{fig:fidelity}.


\subsection{Optimization and Output}

A circuit that clears the fidelity bar is not necessarily minimal
in gate count or depth \cite{fosel}. It therefore passes through an optimization
stage that strips redundant gates and reduces depth whereevr it is 
possible. The reduced circuit is again simulated and its fidelity
checked again; if fidelity goes below the required threshold,
optimization is applied again, more conservatively, until the
threshold is restored. The resulting output package includes the
optimized circuit, its gate sequence, gate count, circuit depth,
fidelity score, and a circuit diagram, as outlined in
Fig.~\ref{fig:circuitopt}.


\subsection{End-to-End Flow}

The complete pipeline runs from user input through target-state
preparation, the RL environment, the DQN/PPO agent, Qiskit
simulation, fidelity evaluation, reward computation, and policy
update, and ends with circuit optimization and a final output that
pairs the optimized circuit with its performance metrics.
Fig.~\ref{fig:dataflow} traces this full data flow, including the
replay buffer, model checkpoint, and training-log stores that
persist across the training loop.

%
%

\clearpage
\onecolumn


\begin{center}

\includegraphics[
    width=0.99\textwidth,
    height=4.7in,
]{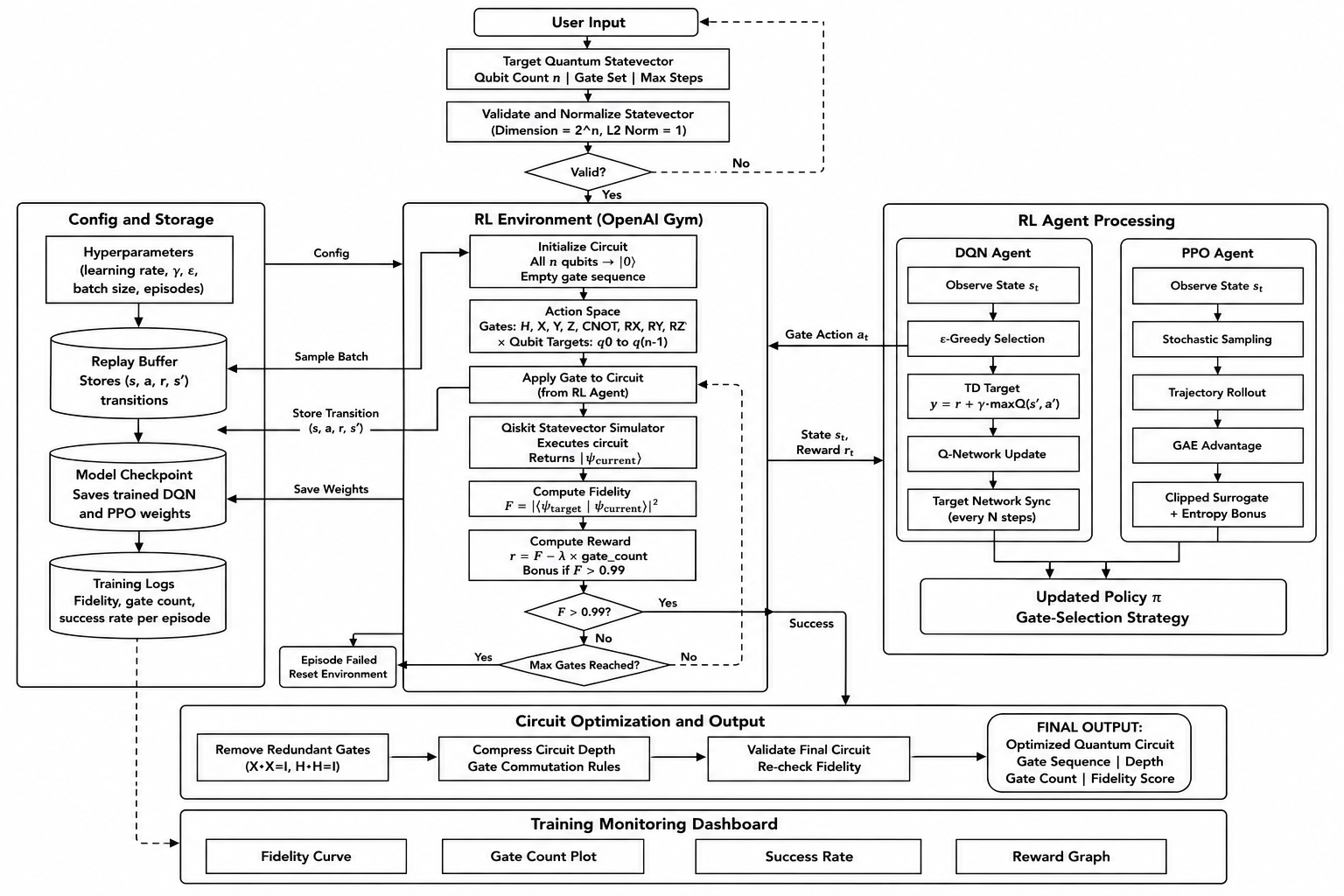}

\captionof{figure}{System architecture of TATVA, showing the interaction between the RL environment, the DQN/PPO agent-processing block, the configuration and storage layer, and the circuit-optimization stage that produces the final output.}

\label{fig:system}

\end{center}

\vspace{0.15in}


\noindent
\begin{minipage}[t]{0.485\textwidth}
\centering

\includegraphics[
    width=0.99\linewidth,
    height=4in,
]{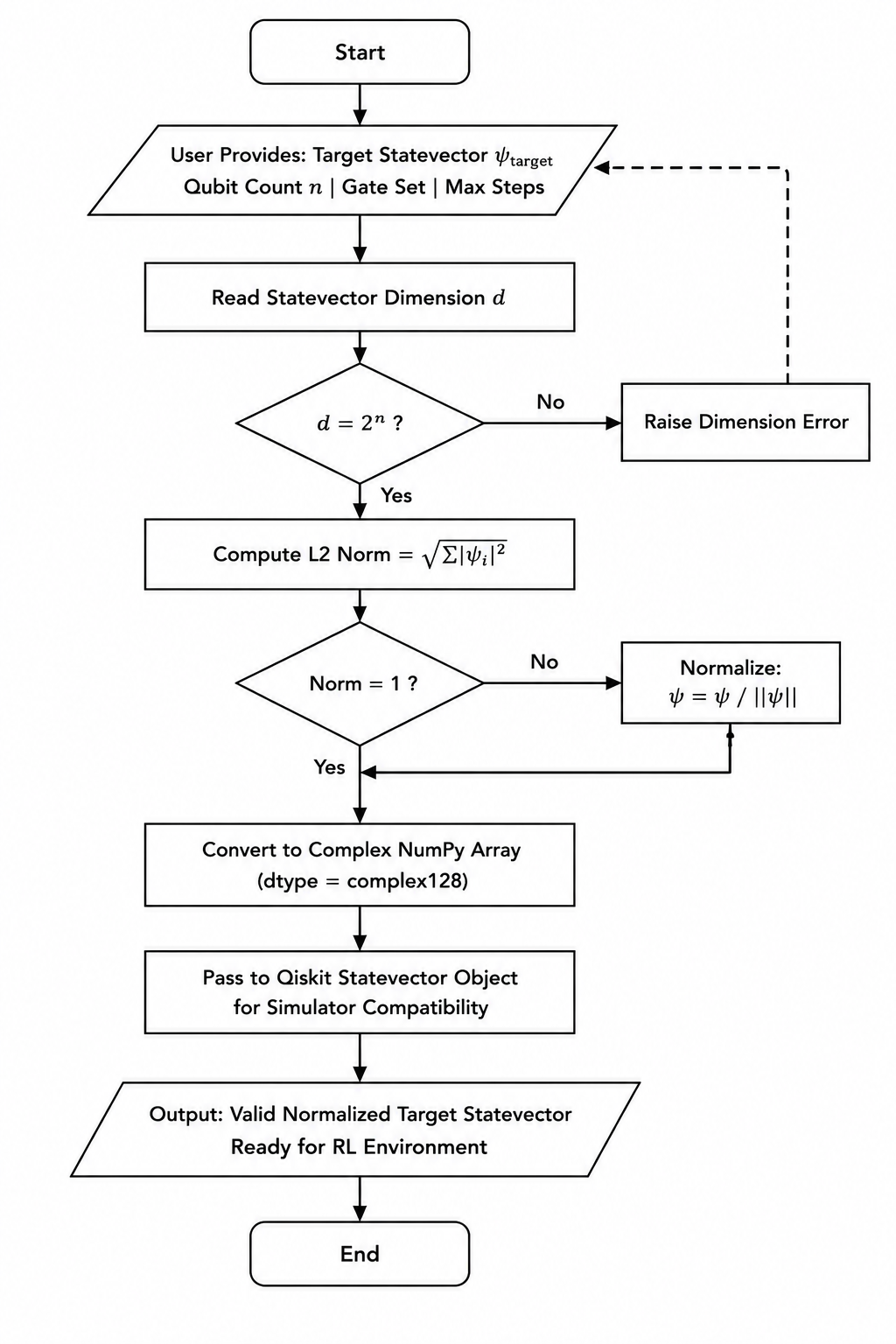}

\captionof{figure}{Input and target-state preparation layer: dimension validation, L2 normalization, and conversion to a Qiskit \texttt{Statevector} object.}

\label{fig:target}

\end{minipage}
\hfill
\begin{minipage}[t]{0.485\textwidth}
\centering

\includegraphics[
    width=0.99\linewidth,
    height=4in,
]{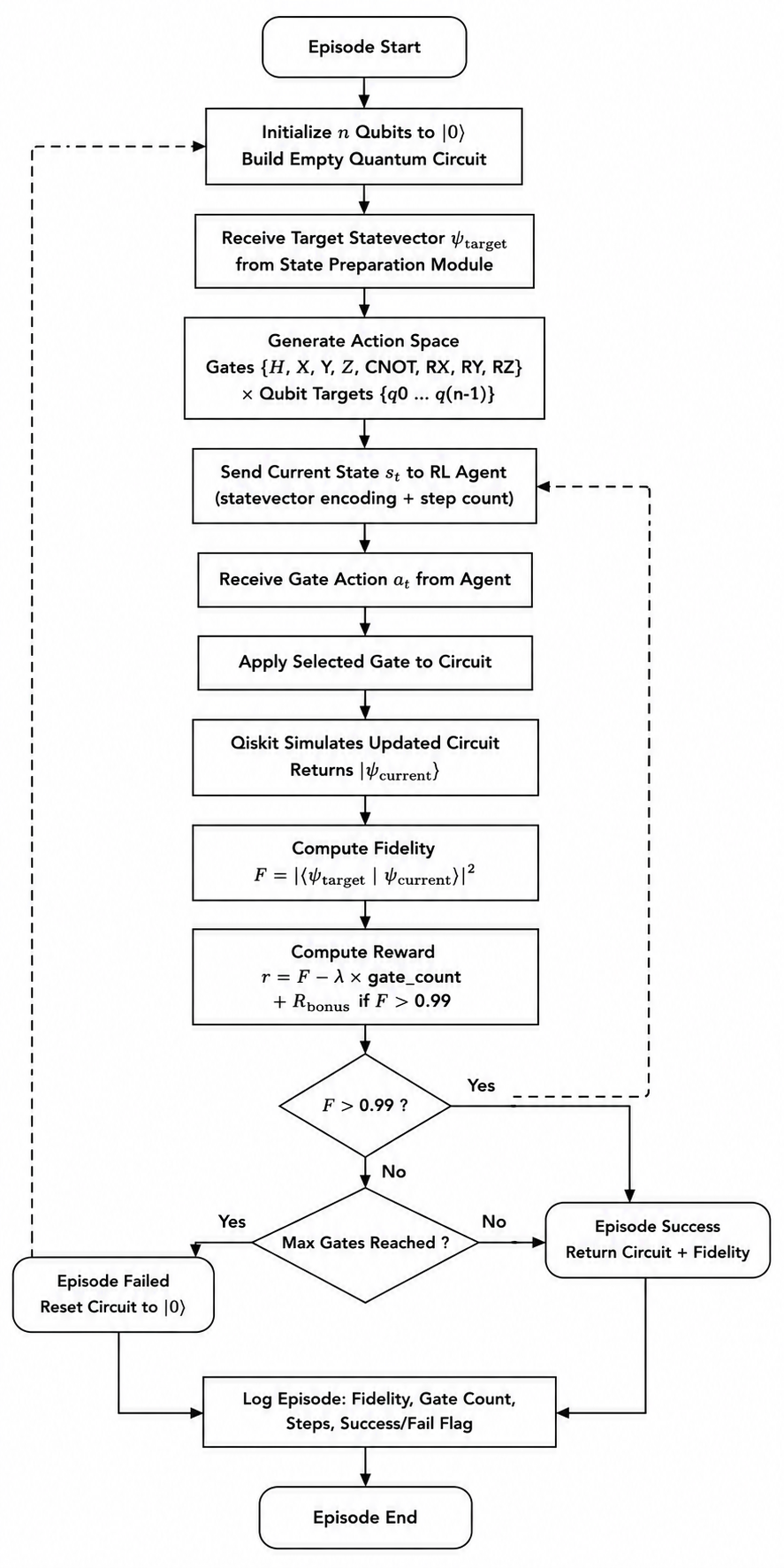}

\captionof{figure}{Reinforcement Learning environment workflow, from episode start through gate application, fidelity/reward computation, and episode logging.}

\label{fig:rlenv}

\end{minipage}

\vspace{0.15in}


\begin{center}

\includegraphics[
    width=0.95\textwidth,
    height=4.5in
]{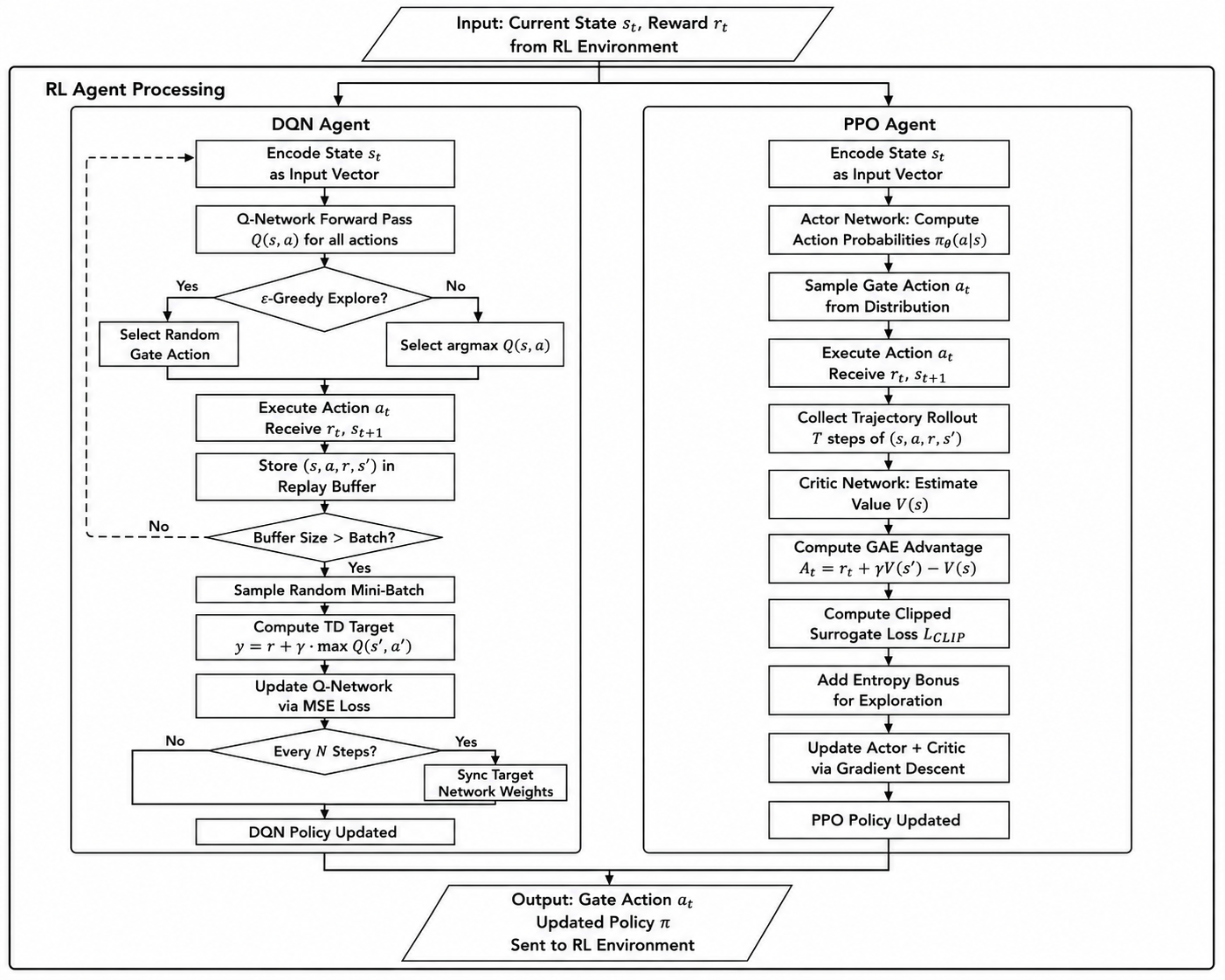}

\captionof{figure}{DQN and PPO agent workflow, run in parallel on the same environment state and reward signal, each producing an updated gate-selection policy.}

\label{fig:dqnppo}

\end{center}

\vspace{0.15in}


\noindent
\begin{center}

\rotatebox{270}{%
\includegraphics[
    width=4in,
    height=1\textwidth
]{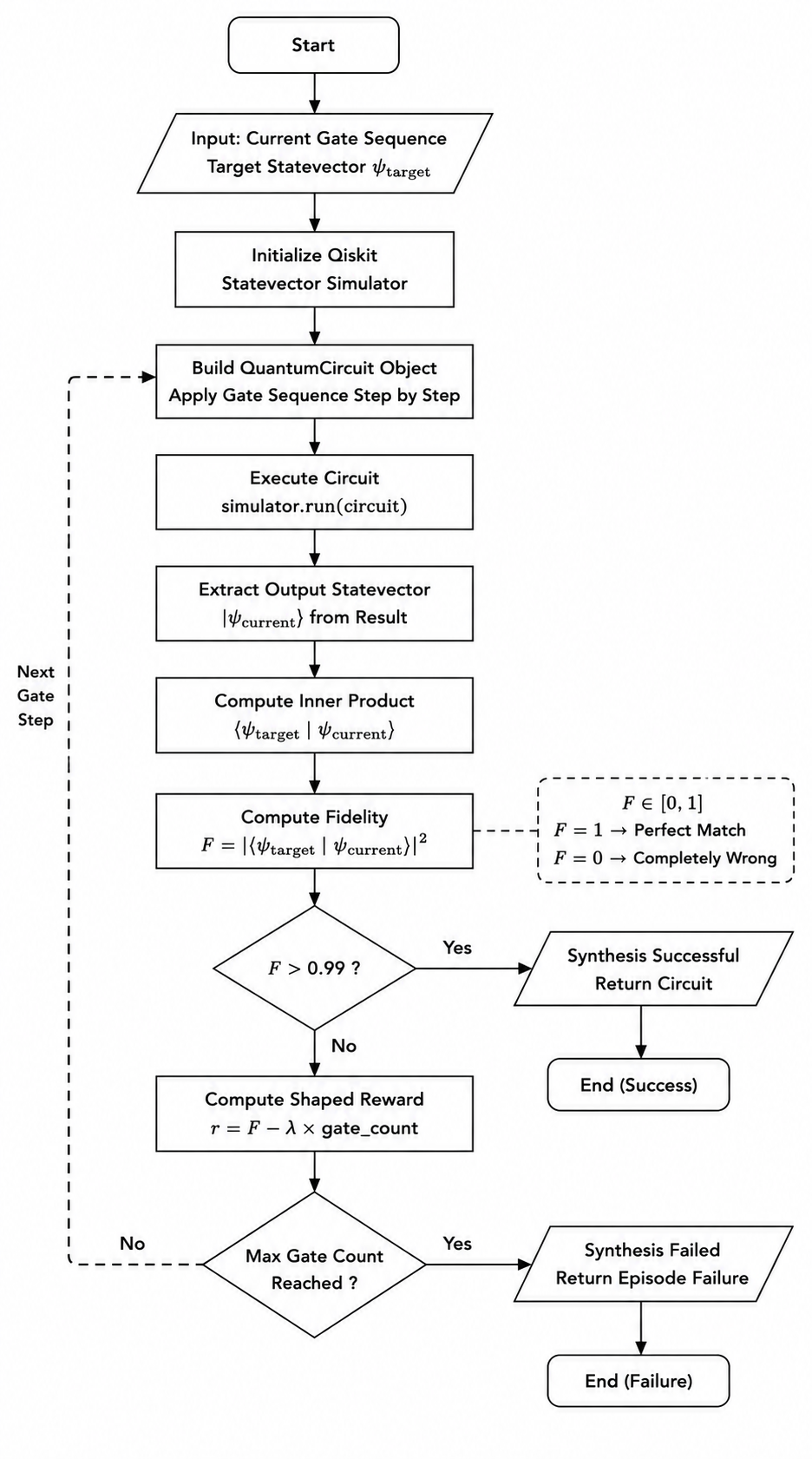}%
}

\captionof{figure}{Quantum simulation and fidelity-evaluation flow: circuit execution, fidelity computation, and the success/continue decision at $F>0.999999$.}

\label{fig:fidelity}

\end{center}

\vspace{0.15in}


\begin{center}

\rotatebox{270}{%
\includegraphics[
    width=4.9in,
    height=0.92\textwidth
]{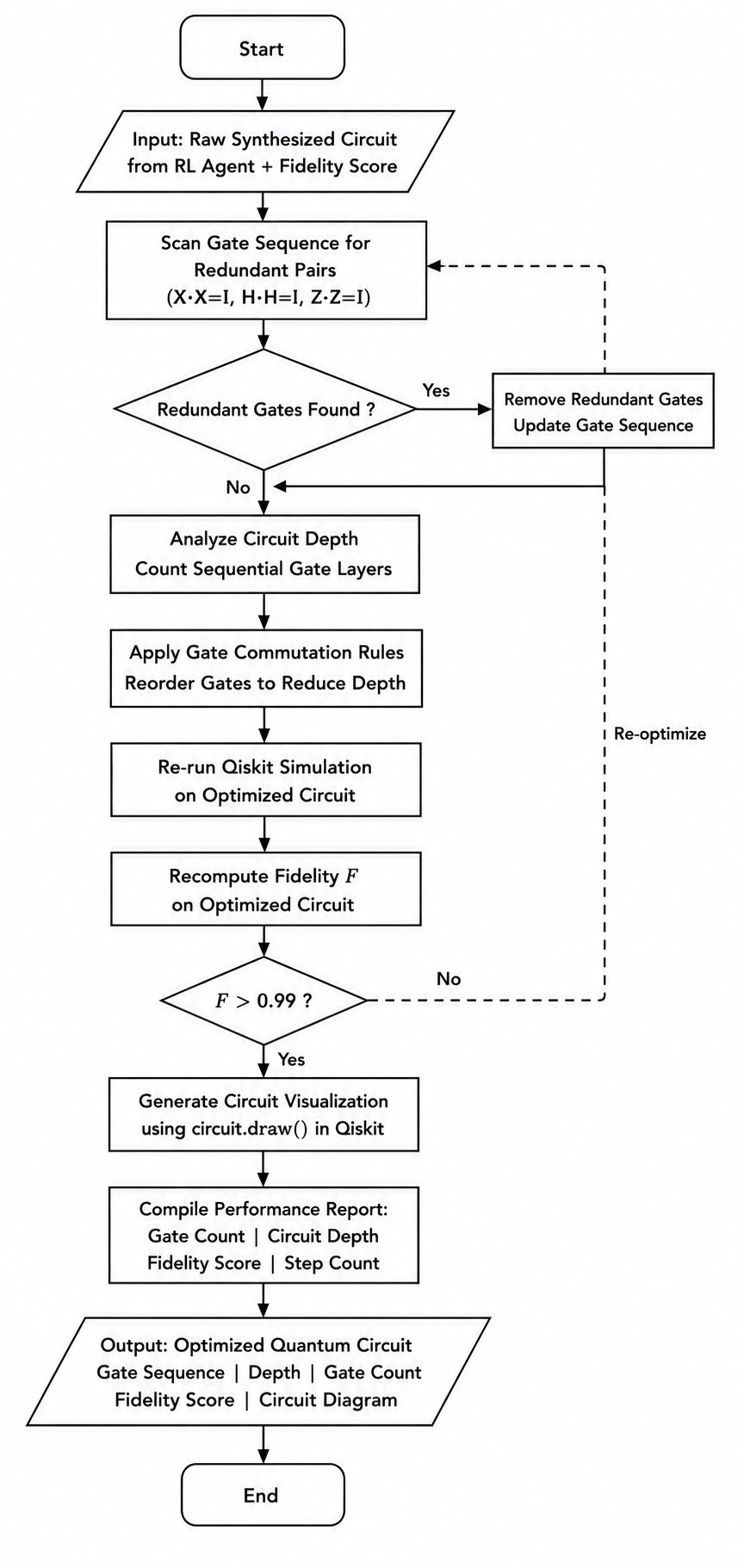}%
}

\captionof{figure}{Circuit-optimization and output-generation workflow: redundant-gate removal, depth compression, fidelity re-check, and final packaging.}

\label{fig:circuitopt}

\end{center}

\vspace{0.15in}


\begin{center}

\includegraphics[
    width=0.98\textwidth,
    height=4.0in
]{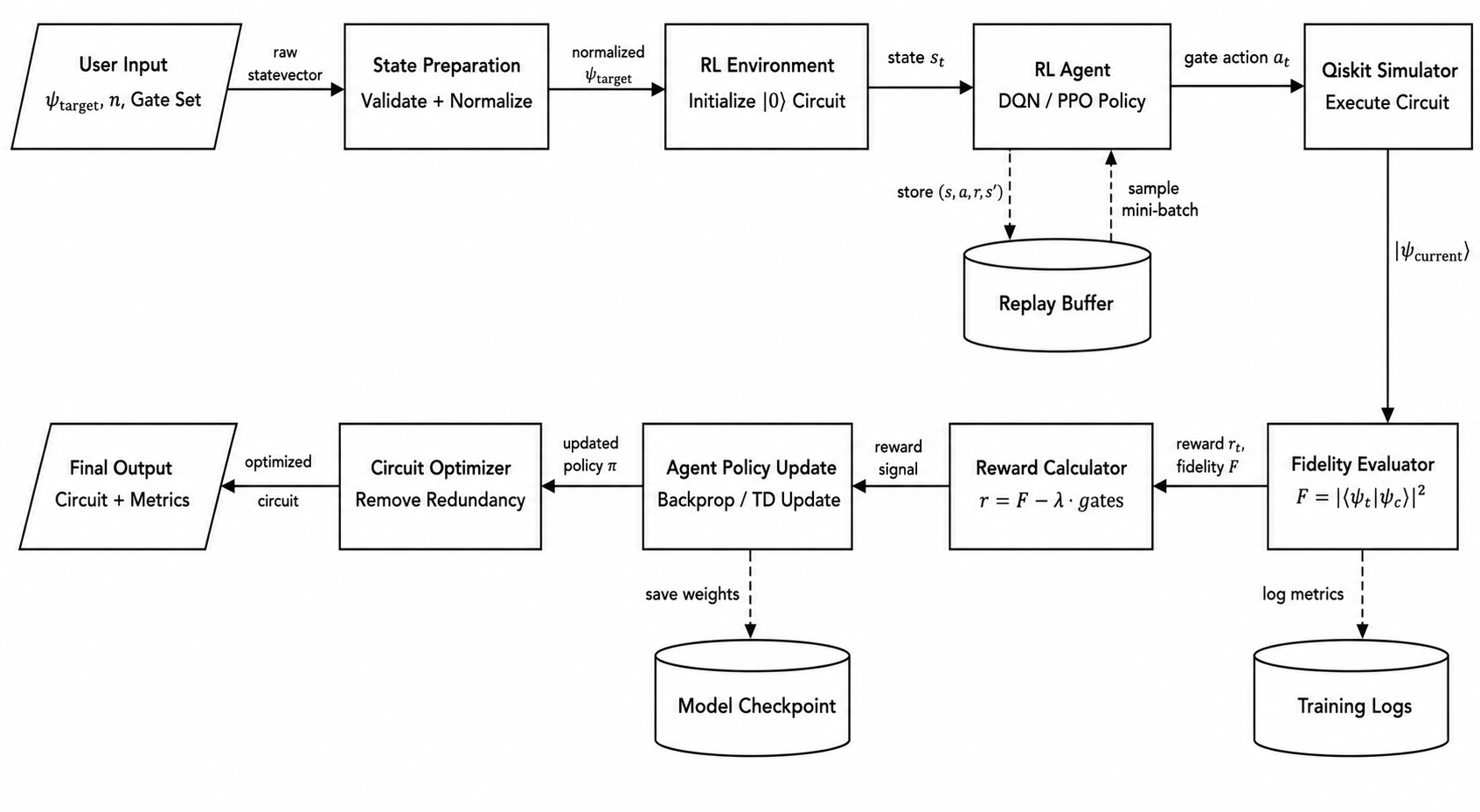}

\captionof{figure}{End-to-end data flow of TATVA, from user-supplied target statevector to the final optimized circuit and its performance metrics.}

\label{fig:dataflow}

\end{center}

\vspace{0.15in}


\clearpage
\twocolumn


\section{RESEARCH AND EVALUATION DESIGN}

The evaluation is built around a single question: can TATVA
consistently produce high-fidelity circuits without paying for that
fidelity in excessive gate count or circuit depth?

\subsection{Target State Categories}

Training and evaluation draw four categories of quantum states which are basis states, superposition states, Bell states, and randomly
sampled Haar states spanning systems of one to five qubits. Previous
reinforcement learning approaches to quantum circuit synthesis have
also considered different target-state configurations and benchmark
quantum states for evaluating circuit synthesis performance
\cite{liu}.

\subsection{Evaluation Procedure}

Each evaluation run starts from $|0\rangle^{\otimes n}$. The agent
appends gates one at a time, with Qiskit re-simulating and recalculating 
fidelity after each addition, until either F $>$ 0.999999 is reached or the gate budget runs out. Beyond measuring performance on the states seen during training, the framework is
also tested against a held-out set of target states which were never
encountered during training. This is a check on generalization rather
than memorization. Fig.~\ref{fig:training} lays out this training
procedure where an inner loop repeats at each gate selection
step until the fidelity threshold or gate budget is reached, while
an outer loop repeats across episodes until the success rate
stabilizes, at which point the final agent weights are
checkpointed.

\subsection{Performance Metrics}

Five metrics anchor the evaluation:

\noindent
\textbf{Fidelity}\hspace{0.35cm}: how close the synthesized state comes to the
target state.

\noindent
\textbf{Success Rate}\hspace{0.10cm}: the share of target states for which
$F > 0.999999$ is achieved.

\noindent
\textbf{Gate Count}\hspace{0.23cm}: the number of gates in the final circuit
\cite{fosel}.

\noindent
\textbf{Circuit Depth}\hspace{0.05cm}: the number of sequential gate layers the
circuit requires \cite{fosel}.

\noindent
\textbf{Generalization}\hspace{0.02cm}: how the trained policy performs on
target states excluded from training.

\subsection{Comparison with Existing Methods}

TATVA is evaluated against three classical baselines which are random
search, greedy search, and Solovay-Kitaev decomposition
\cite{dawson2006}, to compare the quality and efficiency of the generated circuits.
Along with these comparisons, TATVA simultaneously executes both
DQN and PPO models for the given target state. Each model explores
different gate selection decisions and produces a candidate circuit
for the same input. The resulting circuits are then evaluated based
on their achieved fidelity and gate count, and the circuit that
provides the highest fidelity with the most efficient gate count is
selected as the final output. This allows TATVA to make use of both
learning strategies rather than depending on a single RL model.

Prior RL-based work in this area has shown progress in quantum
circuit optimization and state preparation
\cite{fosel,zhang,moro,liu,keren},
but many approaches either involve substantial computational cost,
focus on a limited range of target states, or optimize circuits that
have already been designed 
TATVA instead uses both DQN and PPO from the same target-state input and selects the resulting circuit according to fidelity and circuit efficiency.


\begin{strip}
\centering
\includegraphics[
    width=1\textwidth,
    height=4.2in
]{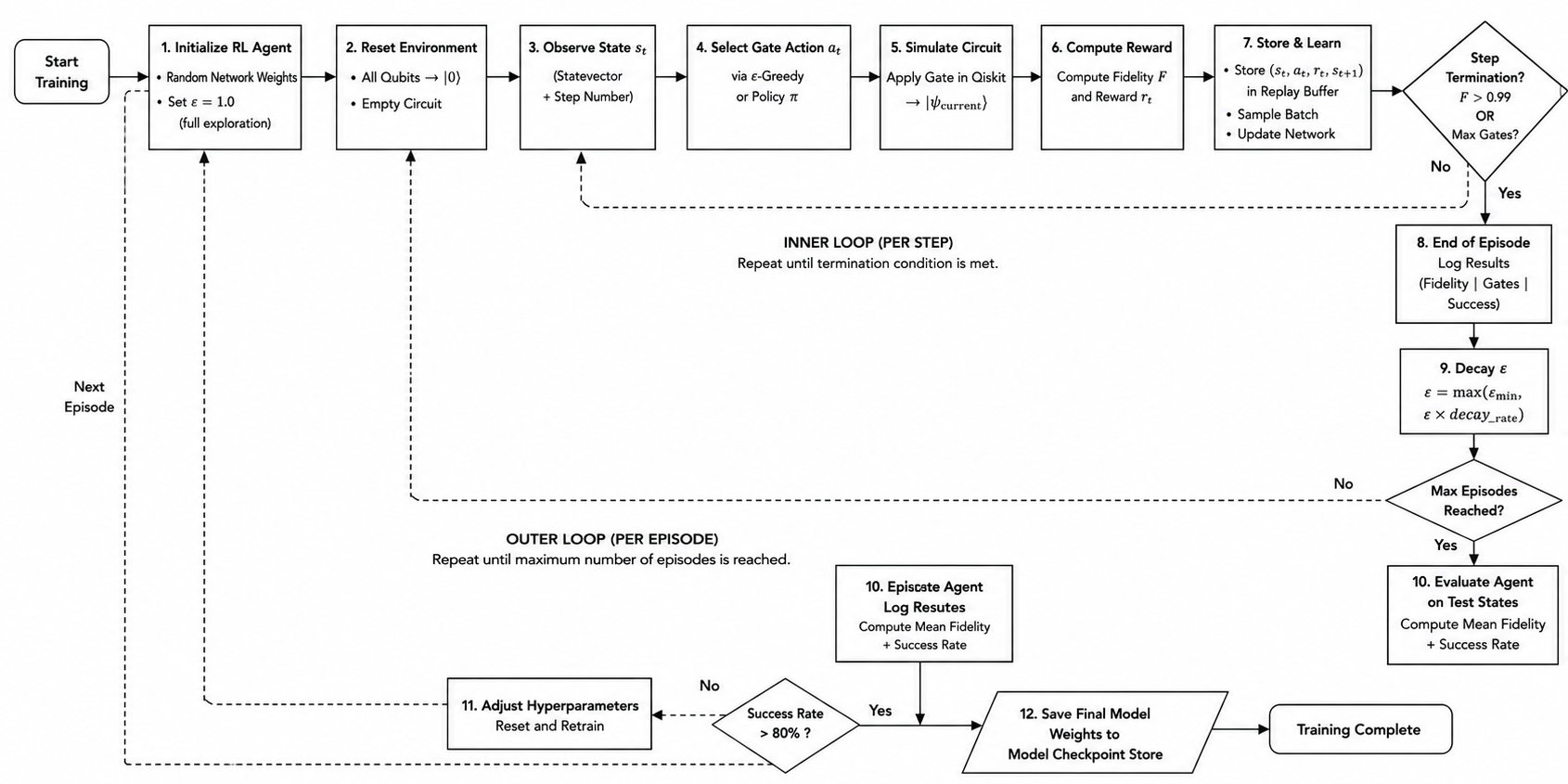}
\captionof{figure}{Training workflow of TATVA, showing the inner per-step loop and the outer per-episode loop that governs epsilon decay, evaluation on held-out states, and final checkpointing.}
\label{fig:training}
\end{strip}


\section{RESULTS AND DISCUSSION}

The TATVA framework was evaluated on one to five-qubit systems,
covering Hilbert spaces from $\mathbb{C}^{2}$ to $\mathbb{C}^{32}$.
The tests included benchmark states such as Bell, GHZ, and W states,
along with arbitrary Haar-random statevectors. The evaluation
focused on four metrics: fidelity ($F$), total gate count
($N_g$), CNOT count ($N_{\text{CX}}$), and circuit depth ($D$).

\begin{table}[H]
\centering

\normalsize
\setlength{\tabcolsep}{5pt}
\renewcommand{\arraystretch}{1.15}

\begin{tabular}{cccccc}
\hline
\textbf{$N$} &
\textbf{Phase} &
\textbf{$F$} &
\textbf{$N_g$} &
\textbf{$N_{\text{CX}}$} &
\textbf{$D$}
\\
\hline

\multirow{2}{*}{1}
& Pre
& 0.999999
& 12
& 0
& 12
\\
& Post
& \textbf{1.000000}
& \textbf{1--3}
& \textbf{0}
& \textbf{1--3}
\\
\hline

\multirow{2}{*}{2}
& Pre
& 0.999999
& 18
& 3
& 15
\\
& Post
& \textbf{1.000000}
& \textbf{4}
& \textbf{1}
& \textbf{4}
\\
\hline

\multirow{2}{*}{3}
& Pre
& 0.999999
& 38
& 8
& 32
\\
& Post
& \textbf{1.000000}
& \textbf{7}
& \textbf{2}
& \textbf{7}
\\
\hline

\multirow{2}{*}{4}
& Pre
& 0.999999
& 96
& 24
& 84
\\
& Post
& \textbf{1.000000}
& \textbf{47}
& \textbf{12}
& \textbf{42}
\\
\hline

\multirow{2}{*}{5}
& Pre
& 0.999999
& 240
& 58
& 214
\\
& Post
& \textbf{1.000000}
& \textbf{114}
& \textbf{26}
& \textbf{106}
\\
\hline

\end{tabular}

\caption{Empirical synthesis and optimization performance across
one- to five-qubit systems.}
\label{tab:results}

\end{table}

\subsection{High-Fidelity Synthesis}
TATVA consistently achieved the target fidelity threshold of
$F \geq 0.999999$ across the one to five-qubit test range.
After continuous L-BFGS-B parameter optimization, the evaluated
circuits reached a numerical fidelity of $F=1.0000000000$,
showing that the synthesized states matched the target states
within floating-point precision.


\subsection{Autonomous Gate Sequence Discovery}

The DQN and PPO agents selected gates autonomously without
hardcoded circuit templates or manual intervention. Single-qubit
states were mainly prepared using single-qubit rotations, while
multi-qubit states led the agents to select entangling operations
such as CNOT, CZ, and SWAP. For example, the two-qubit Bell state
was synthesized using the expected entangling structure with a
single CNOT.


\subsection{Circuit Compaction and Entangling Gate Optimization}

The reward function puts a penalty on unnecessary gates, while the
post processing stage further removes redundant operations and
merges rotation parameters. This produced major reductions
in circuit size:

\begin{itemize}
    \item \textbf{1 qubit:} 12 gates were reduced to 1--3 gates,
    giving a $75.0\%$--$91.7\%$ reduction.

    \item \textbf{2 qubits:} 18 gates and 3 CNOTs were reduced to
    4 gates and 1 CNOT, giving a $77.8\%$ gate reduction.

    \item \textbf{3 qubits:} the GHZ circuit was reduced from
    38 to 7 gates and from 32 to 7 depth levels.

    \item \textbf{4 and 5 qubits:} gate counts were reduced from
    96 to 47 and from 240 to 114, corresponding to reductions of
    $51.0\%$ and $52.5\%$, respectively.
\end{itemize}

The reduction in CNOT operations is particularly important for
NISQ systems, where two-qubit gates generally introduce greater
execution overhead and error.


\begin{strip}
\centering

\includegraphics[
    width=1\textwidth,
    height=4.15in
]{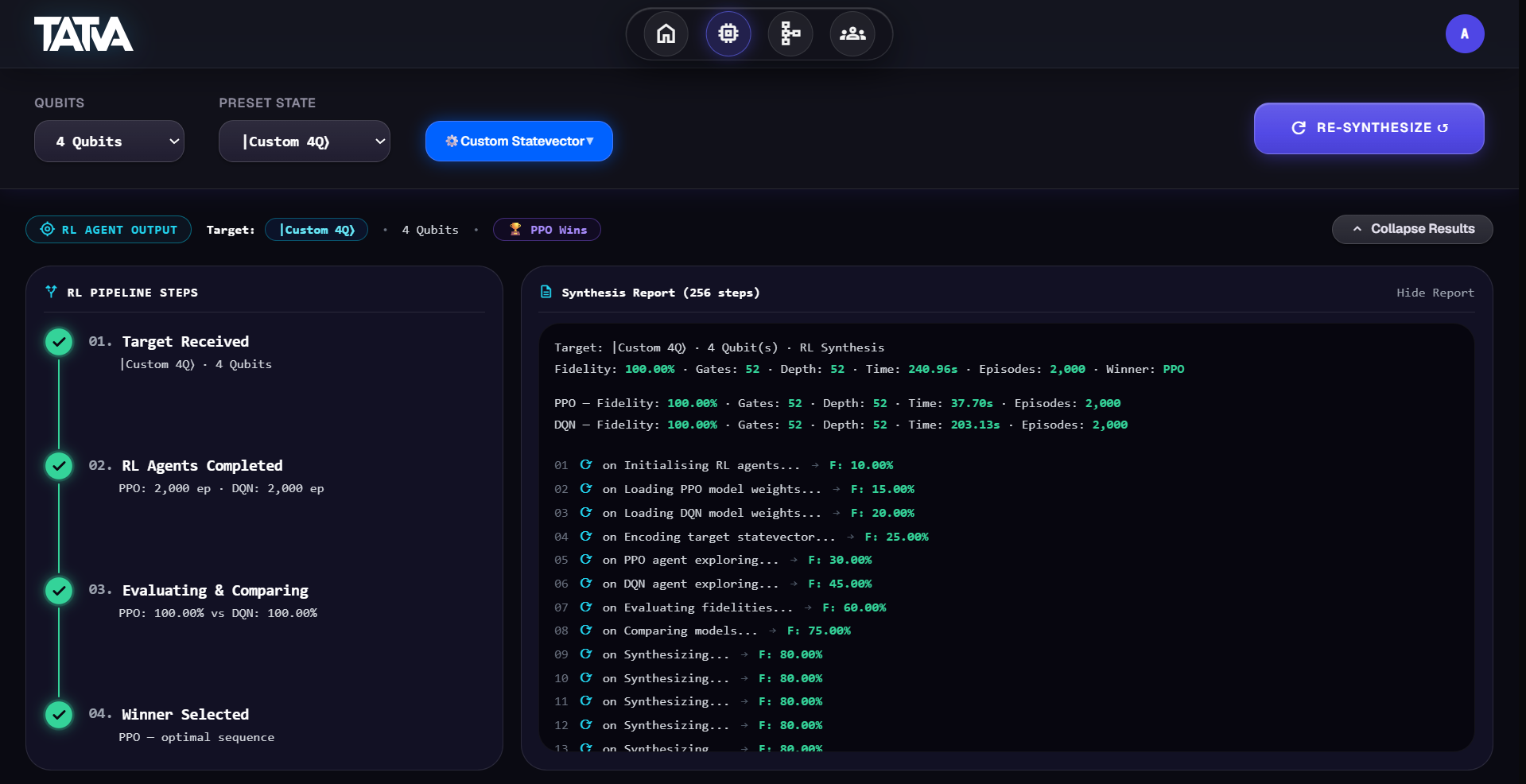}

\captionof{figure}{TATVA platform interface showing the quantum circuit synthesis workflow.}
\label{fig:ss1}

\vspace{0.20in}

\includegraphics[
    width=1\textwidth,
    height=4.15in
]{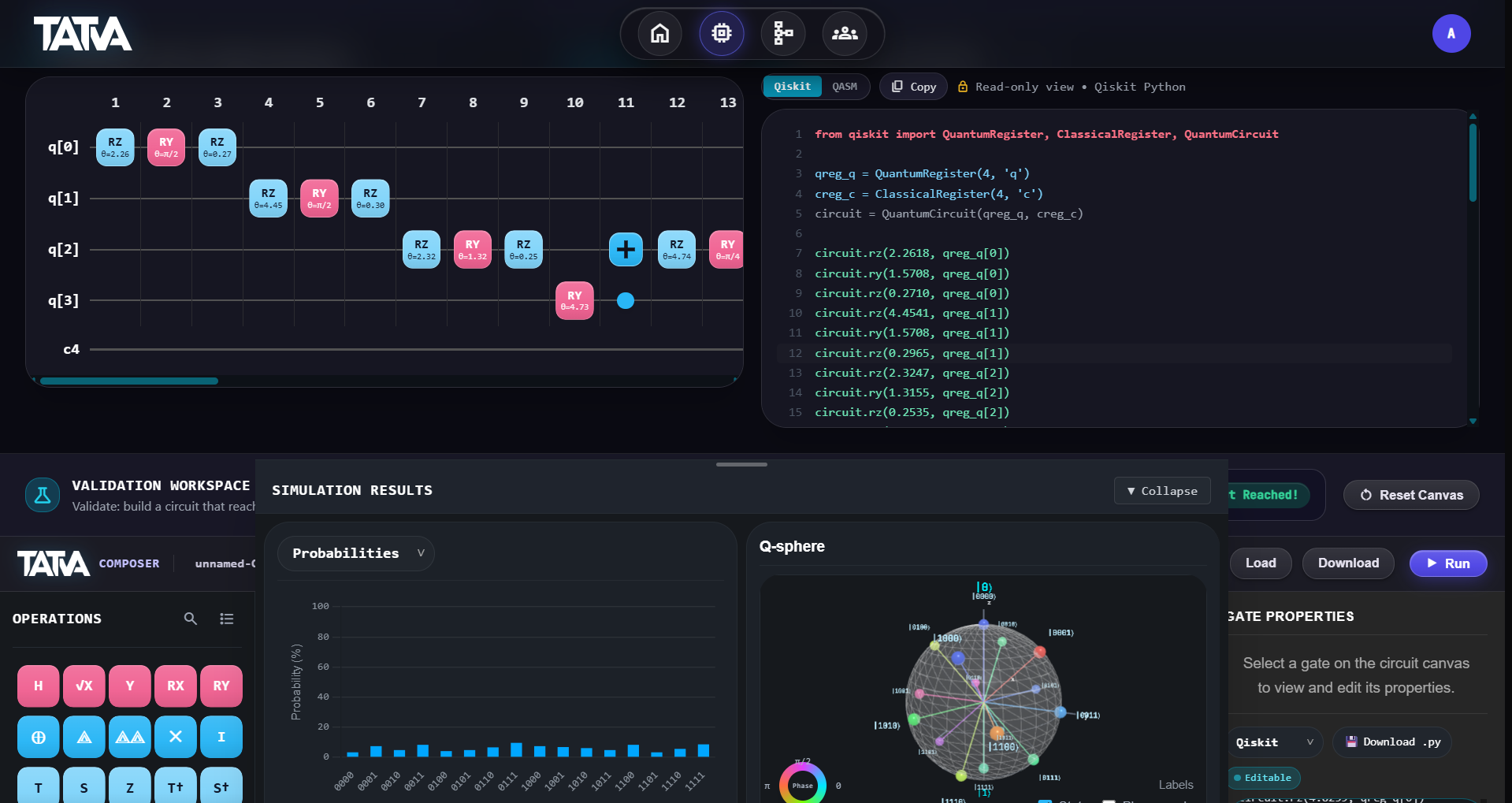}

\captionof{figure}{TATVA platform interface displaying the synthesized circuit and its evaluation results.}
\label{fig:ss2}

\vspace{0.02in}
\end{strip}


\subsection{Generalization Beyond Training}

The trained policies were also tested on unseen Haar-random
statevectors. The results indicate that the agents were able to
apply learned gate selection strategies beyond the states used
during training. On unseen four and five qubit states, the models
continued to generate candidate circuits that could be refined by
the continuous optimization stage.


\subsection{Benchmarking Against Classical Methods}

TATVA was compared with classical approaches including unoptimized
state-preparation routines and Solovay--Kitaev decomposition.
While these methods can achieve accurate state preparation, they
may produce deeper circuits with additional operations. By combining
RL-based gate selection with continuous parameter optimization,
TATVA reduced CNOT usage and circuit depth while maintaining the
target fidelity. This highlights the benefit of combining learned
gate selection with circuit-level optimization.


\section{APPLICATIONS}

The capabilities demonstrated by TATVA extend to several important
areas of quantum computing and quantum information processing, where
accurate and compact quantum state preparation is required.

\textbf{Quantum Machine Learning (QML) State Encoding:}
A major challenge in QML is amplitude encoding, where high-dimensional
classical feature vectors are loaded into quantum state vectors
$|\psi(x)\rangle$. Conventional state preparation circuits can grow
rapidly in depth as the problem size increases. By generating compact
state preparation circuits with verified fidelity
$F \geq 0.999999$, TATVA provides an automated approach for
constructing data encoding circuits that can be used before
executing quantum neural networks (QNNs) or quantum kernel methods.

\begingroup\raggedright
\textbf{Variational Quantum Algorithm (VQA) Initialization:}
Algorithms such as the Variational Quantum Eigensolver (VQE)\allowbreak{} and the
Quantum Approximate Optimization Algorithm (QAOA) often require the
system to start from a specific reference state, such as a
Hartree--Fock state or another problem-specific state. TATVA can
automatically construct compact initial state preparation circuits
and trial circuit structures, helping reduce unnecessary CNOT
operations before variational parameter optimization on NISQ
hardware.
\endgroup

\textbf{Logical State Preparation for Quantum Error Correction:}
Fault-tolerant quantum computing requires the preparation of specific
logical states, including stabilizer-code states used in surface and
color codes, as well as non-stabilizer states such as the magic state
$|T\rangle$. The reinforcement learning approach used in TATVA can
be extended to search for state-preparation circuits for logical
states while considering constraints such as available gate sets and
hardware connectivity.

\textbf{NISQ Hardware Deployment and Transpilation:}
Circuits synthesized by TATVA can be represented using Qiskit and
exported to formats such as OpenQASM for further transpilation and
execution. Since the optimization stage focuses on reducing
two-qubit gate count ($N_{\text{CX}}$) and circuit depth ($D$), the
resulting circuits are better suited for execution on NISQ devices,
including superconducting and trapped-ion quantum processors. Such
reductions can help limit the impact of gate errors and decoherence
during hardware execution.


\section{CONCLUSION}

TATVA presents a reinforcement-learning-based framework for quantum
circuit synthesis, formulating state preparation as a sequential
decision problem in which gates are selected incrementally according
to fidelity and circuit cost. The framework combines target-state
preparation, a custom reinforcement learning environment, DQN and
PPO agents, Qiskit-based simulation, fidelity evaluation, and a
post-hoc circuit optimization stage.

The framework was evaluated on one- to five-qubit systems, covering
basis, superposition, Bell, GHZ, W, and Haar-random target states.
The results show that the synthesized circuits consistently reached
the required fidelity of $0.999999$, while post-hoc optimization
improved the fidelity to $1.0000000000$ in the evaluated cases.
The optimization stage also reduced gate count, CNOT count, and
circuit depth, with the reduction becoming particularly important
for multi-qubit circuits. These results demonstrate that combining
reinforcement-learning-based gate selection with circuit-level
optimization can produce accurate and more compact state-preparation
circuits.

An important aspect of TATVA is the use of both DQN and PPO for the
same target-state input, allowing alternative circuit-generation
strategies to be evaluated using fidelity and circuit efficiency.
The evaluation on randomly sampled Haar states also provides a basis
for studying the ability of the learned policies to generalize beyond
the states used during training.

Future work will focus on extending the framework to larger qubit
systems and broader gate sets, incorporating noise-aware simulation
and hardware-specific constraints, and evaluating the synthesized
circuits on physical quantum devices. These extensions can further
investigate the scalability and practical applicability of
reinforcement-learning-based quantum circuit synthesis.



\section{AUTHOR CONTACT INFORMATION}

The authors' e-mail addresses are as follows:

\noindent
\textbf{Om Bhamare:} bhamareom5@gmail.com

\noindent
\textbf{Aryan Jadhav:} aryanjadhavart2005@gmail.com

\noindent
\textbf{Manas Shinde:} manashshinde@gmail.com

\noindent
\textbf{Prithvi Shinde:} shindeprithvi43@gmail.com

\noindent
\textbf{Dr. Uday Wad:} uday.wad@gmail.com

\noindent
\textbf{Dr. Snehal Kamalapur:} smkamalapur@kkwagh.edu.in

\noindent
\textbf{Dr. S. S. Bhandare:} ssbhandare@kkwagh.edu.in

\end{document}